\documentclass[a4paper,11pt]{article}
\usepackage[utf8]{inputenc}
\usepackage{graphicx}
\usepackage{authblk}
\usepackage{appendix}
\usepackage{epsfig,amssymb,latexsym,amsmath,pifont}
\newcommand{\p}{\varphi}
\newcommand{\e}{\varepsilon}

\newcommand{\om}{\omega}
\author[1]{Aladin Crnki\'c}
\author[2]{Janez Povh}
\author[3]{Vladimir Ja\'cimovi\'c}
\author[4,5]{Zoran Levnaji\'c\footnote{To whom the correspondence should be addressed: \texttt{zoran.levnajic@fis.unm.si}}}

\affil[1]{Faculty of Technical Engineering, University of Biha\'c, Ljubijanki\'ceva, bb., 77000 Biha\'c, Bosnia and Herzegovina} 
\affil[2]{Faculty of Mechanical Engineering, University of Ljubljana, A\v sker\v ceva cesta 6, 1000 Ljubljana, Slovenia}
\affil[3]{Faculty of Natural Sciences and Mathematics, University of Montenegro, Cetinjski put, bb., 81000 Podgorica, Montenegro}
\affil[4]{Complex systems and Data science Lab, Faculty of information studies in Novo mesto, Ljubljanska cesta 31A, 8000 Novo mesto, Slovenia}
\affil[5]{Department of Knowledge Technologies, Jo\v zef Stefan Institute, Jamova cesta 39, 1000 Ljubljana, Slovenia}

\begin{document}

\title{Collective dynamics of phase-repulsive oscillators solves graph coloring problem}

\date{}
\maketitle

\begin{abstract}
We show how to couple phase-oscillators on a graph so that collective dynamics `searches' for the coloring of that graph as it relaxes towards the dynamical equilibrium. This translates a combinatorial optimization problem (graph coloring) into a functional optimization problem (finding and evaluating the global minimum of dynamical non-equilibrium potential, done by the natural system's evolution). Using a sample of graphs we show that our method can serve as a viable alternative to the traditional combinatorial algorithms. Moreover, we show that with the same computational cost our method efficiently solves the harder problem of improper coloring of weighed graphs.
\end{abstract}


\paragraph{\bf Despite the explosion of modern computing power, problems of combinatorial optimizations remain a formidable algorithmic challenge. Drawing upon the theory of collective dynamics of phase-oscillators on graphs (networks), we show how to convert a combinatorial optimization problem (graph coloring) into a functional minimization problem (finding and evaluating the global minimum of a non-negative function on a bounded domain). While this does not (necessarily) offer a simpler solution to the graph coloring problem, it lend it to the whole new world of well-developed methodologies -- those of functional optimization.}


\section{Introduction}
Problems of combinatorial optimization pervade all walks of science~\cite{nemhauser,papadimitriou}. This is particularly true in modern data-driven era, as analyzing the data becomes much harder than obtaining the data. Combinatorial optimization problems revolve around finding the optimal value of some criteria function on a discrete set. That set is usually very large, typical example being the set of all combinations or variations of some elements in a given order. Classical combinatorial optimization problems include linear and quadratic assignments, bin packing problem, vehicle routing problem, graph partitioning, etc.~\cite{schrijver}. 

The difficulty of these problems is hidden in the size of the discrete set on which optimization is to be done. Most of them are NP-hard, which means that there is no algorithm to solve them that works in polynomial time (unless P=NP). In practice, we can optimally solve such problems only for small or medium size instances, while larger instances can usually be solved only approximately and without any approximation guaranty. Combinatorial optimization community developed over the years an array of exact and approximate methods. The winning strategy is usually a combination of general methods with problem-specific ones, like the Hungarian method for linear assignment problem~\cite{schrijver}. Another common approach is to formulate a combinatorial optimization problem as a (non)linear optimization problem with integer constraints and apply Branch and Bound method. But again, these methods quickly fail for larger instances of the problem. The standard alternative is to use various heuristic algorithms, which often involves utilisation of some general meta-heuristic algorithm, like genetic algorithm, simulated annealing, ant-colonies, etc.~\cite{Baghel2012,korte,woeginger2003}. 

Luckily, many optimization problems can be represented in several different ways~\cite{sanz2016}. While this does not make them easier to solve, it helps to  conceptualize them from a different angle and enables different methodology approach them. Excellent example are quantum algorithms, whose promise for solving diverse optimization problems is nowadays clear~\cite{farhi2001}. Quantum simulations that solve optimization problems typically involve finding the ground state of a suitably designed quantum system~\cite{albash2018,johnson2011}. This refers to a variety of optimization problems~\cite{somma2008,wang2018}, including graph coloring~\cite{gaitan2012,gaitan2014,mcmahon2016,kudo2018}. 

On a different scientific front, closer to statistical physics of complex systems, large efforts went into scrutinizing collective dynamics of oscillatory units on complex networks (graphs). Most often this involves studying synchronization and mechanisms of its emergence in relation to the coupling between the oscillators and the underlying graph topology~\cite{arkady,arenas2008,porter,levnajic2010,luciano2011,jacimovic2018,anna2020}. Frequent paradigm are simple phase-oscillators, such as Kuramoto oscillators, where synchronization emerges via positive coupling~\cite{acebron2005,rodrigues2016,crnkic2017,zankoc2019,pietras2019}. The evolving dynamics makes phases of the coupled oscillators `attract' and eventually synchronize. Much less researched is the opposite paradigm, namely, when the oscillators are coupled negatively, i.e., in the phase-repulsive way~\cite{levnajic2011,levnajic2012,goldstein2015,astakhov2016}. Here, the dynamics pushes the phases of the coupled oscillators away from each other. 

In this paper we propose a new way to represent combinatorial optimization problems focusing on the well-known graph coloring problem (vertex coloring). Specifically, we design a family of interaction functions that couple phase-oscillators on a graph and show that natural evolution of such system -- the search for the stationary equilibrium state -- is equivalent to the search for the solution of the vertex coloring problem for the underlying graph. So, to test whether a given graph is colorable with a given number of colors we run the dynamics of phase-oscillators on that graph, coupled via particular interaction function corresponding to the selected number of colors, and examine the equilibrium stationary state. We demonstrate both analytically and numerically that if the graph is indeed colorable with that many colors, oscillators tend to arrange their phases as colors in a solution of the corresponding graph coloring problem. 

Our work is a step forward in recent efforts to use the collective dynamics in various types of oscillators for solving graph (vertex) coloring  problems~\cite{wahwu,wujiao,novikov,parihar}. This includes both theoretical~\cite{wujiao,novikov} and experimental results relying on electric circuits~\cite{wahwu,parihar}. In particular, researchers in \cite{wujiao} and \cite{novikov} observed that by tuning the coupling strength in classic Kuramoto model one finds interesting clustering of equilibrium phases, which can serve as a heuristic for graph coloring. In opposition, our dynamical model involves not only Kuramoto interaction functions (with only the first harmonic), but an entire family of interaction functions, each function with a specific number of harmonics that depends on how many colors are we testing the graph for. As we show in what follows, this enables a better theoretical justification behind our method.


\section{The graph coloring problem}
We begin be precisely formulating the (vertex) graph coloring problem as the combinatorial optimization problem. Let $G=(V,E)$ be a non-directed graph with vertex set $V$ and edge set $E$. A $K$-coloring of vertices $V$ is a function $\varphi \colon V \rightarrow \{ 1,2,\ldots,K \}$, which maps adjacent vertices into different numbers, i.e., $\varphi(u) \neq \varphi(v)$ for all $uv\in E$. Graph is called $K$-colorable if and only if there exists a $K$-coloring of it vertices. The minimal $K$, such that there exists a $K$-coloring is called \textit{chromatic number} of graph $G$ and is denoted by $\chi(G)$~\cite{chartrand}:
\begin{equation}
    \chi(G):=\min \{K \mid \exists \,K\mbox{-coloring of vertices of G}\}.
\end{equation}
Finding the chromatic number of given graph is referred to as \textit{graph coloring problem}, and is a classical problem of combinatorial optimization. There is a vast amount of theoretical results for this problem on different families of graphs and on the problem's extensions~\cite{jensen11}. Graph coloring serves as a theoretical model for several practical problems, including timetable scheduling \cite{ganguli17} or frequency assignment problem \cite{aardal07}.

The problem can be formulated in the context of optimization theory in several different ways~\cite{Gvozdenovic2008,Govorcin2013,Jabrayilov2018}. Here we present the most basic version, so-called assignment-based ILP model~\cite{Jabrayilov2018}:
\[ 
\begin{array}{rrl}
    \chi(G)~=&\min & \sum_{i=1}^{H}w_i \\
     &\mbox{s.t.}&\sum_{i=1}^Hx_{vi}~=~1,~~\forall v\in V\\
     && x_{ui}+x_{vi}\le w_i,~~\forall uv\in E,~i=1,\ldots,H\\
     && x_{vi},w_i\in\{0,1\},~~\forall v\in V,~i=1,\ldots,H
\end{array}
\]
In this model $H$ is a fixed upper bound for $\chi(G)$ obtained by applying some of the many theoretical upper bounds or heuristic algorithms. Binary variables $x_{vi}$ represent the assignment of vertices to colors. Each vertex must be assigned exactly to one color (the first constraint) while two adjacent vertices can not have the same color, which is controlled by binary $w_i$ (the second constraint).

Graph coloring problem is simple for some special families of graphs. Trivial examples are $K$-partite graphs which are $K$-colorable. For the complete graph on $N$ vertices $K_N$ is trivially $\chi(K_N)=N$. For perfect graphs, the chromatic number equals to the size of the largest clique and this number can be efficiently computed by semi-definite programming. This follows from the definition of perfect graphs and by the famous Sandwich theorem~\cite{knuth1994}, while the Strong perfect graph theorem simply characterises these graphs~\cite{schrijver}. 

As for solving the graph coloring problem, exact algorithms that solve it to optimality start with one of the mathematical programming formulations and try to feed it into appropriate nonlinear programming solver, if the problem is small enough and if the there is a solver available for that formulation~\cite{Eppstein2003}. For larger graph sizes, linear and semi-definite programming relaxations are used within Branch and Bound framework. However, for practical needs, a vast amount of heuristic algorithms is available~\cite{Malaguti2010,lewis}.


\section{The dynamical model that colors the graph}
\begin{figure*}[!htb]
\centering
    \includegraphics[width=0.4\linewidth]{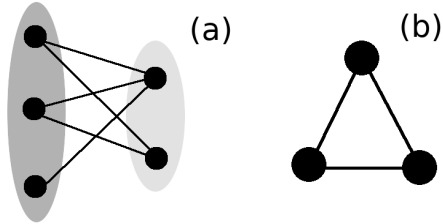} 
    \caption{(a) Illustration of a simple 2-partite graph (network): link connect only nodes in one group with the nodes in the other group. Groups are shown in grey. (b) Illustration of a simple graph that is not 2-partite, but it is 3-partite.}
    \label{fig-1}
\end{figure*}

Coming back to phase-oscillators, we begin explaining the contribution of this paper via illustrative toy-example. Consider a graph with $N$ nodes defined by the adjacency matrix $A$. Phase-oscillators specified by phases $\p_i$ are assigned to its nodes, each with degree $k_i$. Oscillators' frequencies are identical so we set them all to $\om_i=0$ without loss of generality. The coupling strength is $\e=-1$. The equation for this phase-repulsive graph of oscillators reads:
\begin{equation}
\dot \p_i = - (1/k_i) \sum_j A_{ji} \sin (\p_j - \p_i) \; .  
\label{eq-1}
\end{equation}
If our graph had only $N=2$ oscillators (nodes) connected by a link, dynamics would push their phases away from each other to the opposite values. Eventually, the phase difference between them would reach $\pi$, since this is the stable equilibrium for this system~\cite{levnajic2011}. Assume now that our graph is \textit{2-partite}. Dynamics pulls the phases of the connected oscillators apart, i.e., `stretches' the phase differences along each link, trying to reach the maximum stretch value $\pi$. Since the graph is 2-partite, this equilibrium is easily attained. Eventually, oscillators in two groups will have the opposite phase values and the graph will be perfectly anti-synchronized. An illustration of this situation is shown in Fig.~\ref{fig-1}a. 

But now, inverting this problem, Eq.~\ref{eq-1} can be used to check whether a given graph is 2-partite. Once the dynamics of Eq.~\ref{eq-1} reaches equilibrium, one looks at the phase differences along links: if they are all $\pi$ the graph is 2-partite. Actually, consider a graph that is not 2-partite. For simplicity, let it have 3 nodes connected in a triangle, like in Fig.~\ref{fig-1}b. It is a 3-partite graph, since nodes can be grouped in three groups with links going only between the groups (but not in two groups). The dynamics of Eq.~\ref{eq-1} will no longer be able to 'stretch' the phase differences along links to $\pi$. Instead, at least some links will be ``frustrated" or ``squeezed" to a phase difference less than $\pi$. This will happen for any choice of initial phases, indicating that the underlying graph is not 2-partite. 

To articulate this dynamical situation we introduce frustration $f_{ij} = f_{ij}(\p_j - \p_i)$ along the link $i-j$ ($A_{ij} = 1$) as:
\begin{equation}
 f_{ij}(\p_j - \p_i) = 1 + cos(\p_j - \p_i) \; . 
 \label{eq-2}
\end{equation}
Frustration captures how ``squeezed" is a link, $f_{ij}=0$ means that link $i-j$ is not frustrated (phase difference is $\pi$). If we picture the oscillators' interactions as springs, frustration can be thought of as elastic potential energy due to spring being squeezed~\cite{levnajic2011}. In fact, gradient of $f_{ij}$ yields the sinusoidal interaction in Eq.~\ref{eq-1}, so $f$ plays the role of dynamical non-equilibrium potential~\footnote{Frustration can be equivalently defined for phase-attractive case, but it is always zero since for identical oscillators full synchronization is the only equilibrium state. In contrast, for the phase-repulsive case graph topology is intimately related to frustration~\cite{levnajic2011}}. We define the total frustration $F$ of a graph as the sum of frustrations along the individual links:
\begin{equation}
F = \sum_{i,j} A_{ij} f_{ij}(\p_j - \p_i) \; ,
\label{eq-total}
\end{equation}
so that $F=0$ guarantees that $f_{ij}$ is zero along all links. The graph in Fig.~\ref{fig-1}a will always have $F=0$, whereas the one in Fig.~\ref{fig-1}b will always have at least some frustrated links and hence $F>0$. This shows that former graph is 2-partite, while the latter is not. Eq.~\ref{eq-1} can be used to confirm 2-partitness, but not to confirm 3-partitness. 

So, can we develop a dynamical model inspired by Eq.~\ref{eq-1} that could confirm $K$-colorability of a graph for any $K$ (not just for $K=2$)? Such model would offer a dynamical solution to the combinatorial optimization problem~\cite{sanz2016}, finding the chromatic number \textit{dynamically}, just by naturally relaxing to its equilibrium. Note the analogy with searching for the ground state in quantum systems~\cite{kudo2018} (quantum algorithms here serve \textit{only} as inspiration and have no relevance otherwise).

Equipped with above preliminaries, we now present exactly such a model. We are given a non-weighted and non-directed graph and set out to find a dynamical model that checks its $K$-colorability for any $K$. We need the specific coupling functions and frustration function for each $K$. For $K=2$ we have Eq.~\ref{eq-1} and Eq.~\ref{eq-2}: a graph is 2-colorable if for at least one choice of initial phases we obtain $F=0$.

We define the $K$-frustration $f^K_{ij}$ along the link $i-j$ for $K \geq 2$ as:
\begin{equation}
f^K_{ij} (\p_j - \p_i) = 1 + c_K \Big[ (K-1) \cos (\p_j - \p_i) + (K-2) \cos 2(\p_j - \p_i) + \cdots + \cos (K-1)(\p_j - \p_i) \Big] \; . 
\label{eq-Kfrustation}
\end{equation}
Constants $c_K$ are fixed to have $f^K_{ij} (\frac{2\pi}{K})=0$ for every $K$. Each $f^K$ is a trigonometric polynomial of the degree $K-1$. For $K=2$ we have $f^2_{ij} = 1 + c_2 \cos (\p_j - \p_i)$. Requesting $f^2_{ij}(\pi)=0$ we get $c_2=1$ and recover Eq.~\ref{eq-2}. 

We define the total $K$-frustration $F^K$ similarly to Eq.~\ref{eq-total} as the sum of $K$-frustrations along the individual links 
\begin{equation}
F^K  = \sum_{i,j} A_{ij} f^K_{ij} (\p_j - \p_i) \; .  
\end{equation}
$F^K$ is the non-equilibrium potential for the dynamical model behind Eq.~\ref{eq-Kfrustation}. We have:
\begin{equation}
\dot \p_i = - \frac{1}{k_i} \frac{\partial F^K}{\partial \p_i} = - \frac{1}{k_i} \sum_j A_{ij} \frac{\partial}{\partial \p_i} f^K_{ij}  (\p_j - \p_i) =
- \frac{1}{k_i} \sum_j A_{ij} C^K (\p_j - \p_i) \;  . 
\label{eq-Kdynamics}
\end{equation}
For coupling functions $C^K$ we get:
\[ C^K (\p_j - \p_i) = c_K \Big[ (K-1) \sin (\p_j - \p_i) + 2 (K-2) \sin 2(\p_j - \p_i) + \cdots + (K-1) \sin (K-1)(\p_j - \p_i) \Big] \; , \]
where the constants $c_K$ are fixed as before. Since $c_2=1$, Eq.~\ref{eq-Kdynamics} for $K=2$ reduces to the phase-repulsive graph of Kuramoto oscillators Eq.~\ref{eq-1}. 

For higher values of $K$, in addition to the first harmonic like in Eq.~\ref{eq-1}, oscillators are also coupled via higher harmonics. Consider the example of $K=3$. We have:
\begin{equation}
\begin{aligned}
 f^3_{ij} (\p_j - \p_i) = 1 + \frac{4}{3} \cos (\p_j - \p_i) + \frac{2}{3} \cos 2(\p_j - \p_i)  \; , \\ 
 \dot \p_i = - \frac{4/3}{k_i} \sum_j A_{ji} \Big[ \sin (\p_j - \p_i) + \sin 2 (\p_j - \p_i) \Big] \; . 
\end{aligned} 
\end{equation}
For phase differences $\p_j - \p_i = \frac{2\pi}{3}$ or $\p_j - \p_i = \frac{4\pi}{3}$ we get $f^3_{ij}=0$ and hence $\dot \p_i = 0$. In contrast, for all other phase differences we have $\dot \p_i \neq 0$. Hence, above dynamical model looks for equilibrium states where the phase differences along links are integer multiples of $\frac{2\pi}{3}$. Back to the graph from Fig.~\ref{fig-1}b, under dynamics governed by $F^3$ its three nodes will ultimately have three different phase values, equidistant on the circle, and separated by $\frac{2\pi}{3}$. In this situation we have $F^3=0$, indicating that this graph is 3-colorable. Meanwhile, $F=F^2$ will stay positive, since graph is not 2-colorable (every $K$-colorable graph is also $(K+1)$-colorable, but not vice versa). 

In general, \textit{any graph is $K$-colorable if and only if the global minimum of $F^K$ is zero} and the  smallest such $K \geq 2$ is the graph's chromatic number $\chi(G)$. In fact, if for some arrangement of equilibrium phase values around the nodes we find $F^K=0$, then there are exactly $K$ different equilibrium phase values. This means some nodes will have common phase values, but never those that are connected. Instead, phases of the connected nodes will always be integer multiples of $\frac{2\pi}{K}$ apart (if they were not, this would not be an equilibrium state, and $F^K$ would be positive). Identifying $K$ colors as $K$ equilibrium phase values, we have that such graph is $K$-colorable. This is the core result of our paper and it is rigorously proven in the Appendix. 

To confirm that a graph is $K$-colorable one needs to run the dynamical model Eq.~\ref{eq-Kdynamics} from many initial phases and examine the final equilibrium states, searching for the one with $F^K=0$. Simplest way to do this is to track $F^K(t)$ (in addition to several `tricks' that we discuss later). Note that $F^K$ \textit{always has} a global minimum: the question is, if it is zero. Thus, checking the $K$-colorability of a graph is translated into evaluating the global minimum of a real function $F^K$, which is done via natural evolution of the collective dynamics Eq.~\ref{eq-Kdynamics}. We have thus translated a combinatorial into a continuous optimization problem: rather than permuting arrangements of node colorings (combinatorial problem), we run the Eq.~\ref{eq-Kdynamics} by integrating ordinary differential equations (continuous problem). 

However, if $F^K=0$ is not found, this still does not prove that graph is not $K$-colorable. $F^K$ is a trigonometric polynomial whose functional properties depend on (and in fact, capture) the topology of the underlying graph. Therefore, in addition to the global minimum, such function will in general have many local minima where $F^K>0$. Dynamics could easily get ``stuck" in one of them, and confuse it for the global minimum, leading to (potentially false) conclusion that graph is not $K$-colorable. Dynamics can miss the true global minimum for several reasons, e.g. because its basin of attraction is too small. This is a notorious problem in functional optimization.

Furthermore, there are situations where the minimization converges not to a point, but to a critical set of bigger dimensionality, for example a limit cycle. However, this does not change the principle of our method: we look for $F^K=0$, regardless of the dimensionality of the set on which this occurs. In fact, larger the set with $F^K=0$ is, easier is for convergence process to find it. On the other hand, convergence can become stuck more easily in a local minimum of bigger dimensionality (than in a point). Partial remedy for this is in `tricks' that we discuss later.

\paragraph{An important remark.} Once the non-negative function $F^K$ is defined, its local and global minima are what they are (zero or otherwise), regardless of whether our dynamics Eq.~\ref{eq-Kdynamics} finds them or not. So, running the dynamics Eq.~\ref{eq-Kdynamics} \textit{is not the only way} to evaluate the global minimum of $F^K$. There are numerous approaches in functional optimization for evaluating the minima of real functions, such as gradient descent, differential evolution, etc~\cite{sundaram}. In fact, some of them could be more efficient than the natural evolution of Eq.~\ref{eq-Kdynamics}. In other words, once the combinatorial problem of $K$-colorability is converted into the optimization (evaluation) problem, it can be approached either via Eq.~\ref{eq-Kdynamics} or via any of the standard methods in functional optimization.


\section{Simulations}
To illustrate our result, we construct a simple graph with 8 nodes and chromatic number 4. We run the dynamics of Eq.~\ref{eq-Kdynamics} with coupling function $C^4$, whose corresponding frustration is $F^4$. Initial phases are set at random, but instead of picking them from uniform distribution on the circle, we use von Mises distribution~\cite{mardia} (it can roughly be thought of as a Gaussian on the circle, where $\kappa \geq 0$ plays the role of standard deviation, so that uniform distribution is obtained for $\kappa = 0$). The effect of this is that initially all phases are close to each other, so the system is very frustrated, which gives it a bit of ``push" and makes phases diverge more rapidly. 

\begin{figure*}[!hbt]
\centering
    \includegraphics[width=\textwidth]{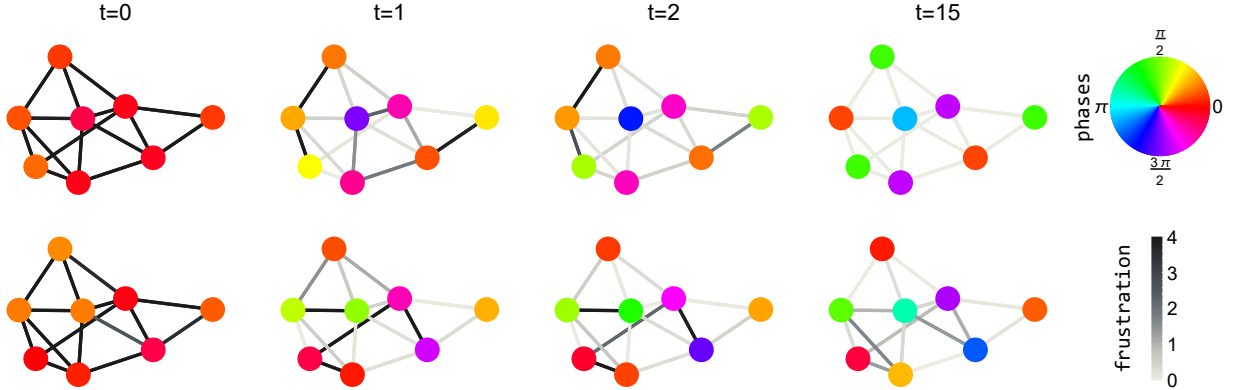} 
    \caption{Two realizations of evolution for illustrative graph with 8 nodes, shown in four time snapshots as indicated above. Both evolutions start from initial phases randomly chosen from von Mises distribution on the circle with $\kappa=10$~\cite{mardia}. The phases of the nodes are shown by colors and frustration of links in gray-scale (see two colorbars on the right). Top panel is an example of dynamics that reaches $F^4=0$, while bottom panel is an example of dynamics that settles in a local minimum with $F^4>0$.} 
    \label{fig-2}
\end{figure*}

In Fig.~\ref{fig-2} we show two realizations of dynamical evolution on the graph via four snapshots (last snapshot for $t=15$ is the equilibrium state). In top panel the dynamics reaches the state with $F^4=0$, which confirms that graph is 4-colorable. There are indeed only 4 different final phase values and no connected pair of nodes is colored the same. In bottom panel the dynamics settles into a state with $F^4>0$, since some links remain frustrated and there are more than 4 different equilibrium phase values. This is a local minimum, in the sense that small perturbations will not ``kick" it out of it. 
This shows how not all runs finish in the global minimum, regardless of it being zero or not. Still, we need only one realization as in top panel to confirm that graph is 4-colorable.  

\begin{figure*}[!hbt]
\centering
    \includegraphics[width=\textwidth]{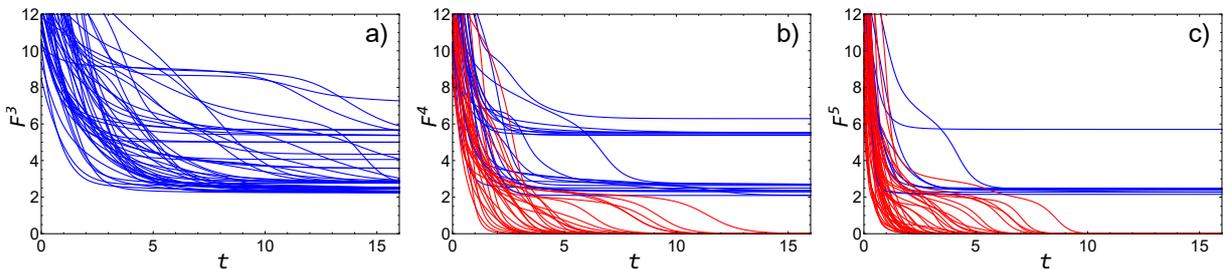}  
    \caption{(a) Plots of $F^3 (t)$ for 50 runs of Eq.~\ref{eq-Kdynamics} with $C^3$ for the same graph as in Fig.~\ref{fig-2}. (b) The same for $C^4$ and $F^4$, two examples on this dynamics are shown in Fig.~\ref{fig-2}. (c) The same for $C^5$ and $F^5$. Plots are shown until $t=15$ for consistency with Fig.~\ref{fig-2}. Red curves are runs that finish in $F^K=0$ and blue curve are runs that finish in $F^K>0$.} 
    \label{fig-3}
\end{figure*}

But what if the basin of attraction for the state $F^4=0$ was much narrower? In that case it could happen that no runs finish in the global minimum. This would falsely lead us to believe that graph is not 4-colorable. Establishing whether some minimum is local or global is a notorious problem in functional optimization. To investigate its repercussions here, we trace $F^K (t)$ starting from 50 initial phases for above graph and report the results in Fig.~\ref{fig-3}. Situation in earlier Fig.~\ref{fig-2} corresponds to  Fig.~\ref{fig-3}b ($K=4$). Only a fraction of runs finish in $F^4=0$ (red curves), and the rest finishes in $F^4>0$ (blue curves). Examples of the former and the latter correspond to the top and the bottom panels in Fig.~\ref{fig-2}, respectively. One needs to run the dynamics from many different initial phases to find one state with $F=0$, assuming there are any. 
To put this in context, we re-do the same runs with $K=3$ are show the plots in Fig.~\ref{fig-3}a. Clearly, none of the runs finish with $F^3=0$, since graph is not 3-colorable. However, the same situation could happen for $K=4$ unless we run enough realizations. In contrast, in Fig.~\ref{fig-3}c we show the same for $K=5$: almost all runs lead to $F^5=0$, since graph is easily colorable with 5 colors. Confirming the colorability for higher $K$ is easier, since $F^K$ touches zero in more points, despite $F^K$ being a higher-order trigonometric polynomial for higher $K$. As we get closer to the minimal $K$ for which the graph is $K$-colorable, less and less initial phases lead to $F^K=0$, since the basin of attraction for states with $F^K=0$ shrinks. That is why pinpointing the chromatic number of a graph is challenging, as illustrated in Fig.~\ref{fig-3}.

We ran many more simulations for graphs of varying size and chromatic numbers. Overall, we found that our simulations correctly identify the chromatic numbers obtained via traditional combinatorial algorithms. However, as graph size grows beyond $N=50$, the number of necessary runs increases rapidly. In fact, the dimensionality of space on which $F^K$ is defined is $N$. As $F^K$ becomes higher and higher dimensional, pinpointing the basin of attraction for $F^K=0$ becomes harder and harder. But still, if the chromatic number is (relatively) low, the corresponding $F^K$ is a low-order trigonometric polynomial, so our dynamical graph coloring works. 

On the other hand, except taking the initial phases from specific distributions, there are other enhancements (or `tricks') one could use. One of them is to introduce kicking in the oscillator dynamics, i.e., make Eq.~\ref{eq-Kdynamics} a stochastic differential equation by adding a noise term on its RHS. The effect of this is that evolution can get kicked out of a local minimum, possibly finding the global minimum later. This is similar to the effect that temperature has in simulated annealing. We confirmed that this enhancement indeed improves the overall performance of our method, but it still does not solve the challenge of large graphs and high chromatic numbers.


\section{Improper coloring of weighted graphs}
Our approach is actually useful for a harder problem known as improper graph coloring~\cite{riste1999,mohar2003}. Here, one looks into colorings with number of colors smaller than graph's chromatic number. Therefore, one allows for some connected nodes to be colored the same, but looks for the arrangement of colors that minimizes the number of such pairs. In this improper sense, any graph is colorable with any $K$ number of colors, but the question is, how to minimize  the number of connected nodes having the same color. If this number is zero for some $K$, the graph is $K$- colorable in the proper sense, and we have $\chi(G)\le K$. Several variations of this problem have been studied in the literature, although not very frequently, with diverse motivations, for example related to scheduling and timetabling problems~\cite{mishra2005}. 

To make our analysis even more general, in this section we face the problem of improper coloring of weighted graphs. In a weighted graph, links (edges) are not all of equal weight (``thickness"), but can be stronger or weaker, modeling the strength of interactions between the nodes. Consider a non-directed weighted graph with $N$ nodes, described by the symmetric weighted adjacency matrix $W_{ij}$. The entry $W_{ij}$ in this matrix denotes the (non-negative) weight of the link $i-j$. Given an integer $K \leq N$, the problem is to color the nodes of this graph in a way to minimize the sum of weights of links that connect nodes that are colored the same. In other words, we want to minimize the sum of $W_{ij}$ such that nodes $i$ and $j$ are colored the same (if $i$ and $j$ are not connected then of course $W_{ij}=0$). Traditionally, improper graph coloring is treated via similar combinatorial algorithms, which in weighted case have an additional layer of complexity.

As before, we assign phase-oscillators of identical frequencies $\om=0$ to graph nodes and consider a tentative coloring number $K$. $K$-frustration $f^K_{ij}$ along $i-j$ remains defined by Eq.~\ref{eq-Kfrustation}, but the total $K$-frustration $F^K_W$ is now a weighted sum of individual link frustrations:
\begin{equation}
    F^K_W  = \sum_{i,j} W_{ij} f^K_{ij} (\p_j - \p_i) \; .
\end{equation}
The dynamical equations are obtained the same way:
\begin{equation}
  \dot \p_i = - \frac{1}{k_i} \frac{\partial F^K_W}{\partial \p_i} = - \frac{1}{k_i} \sum_j W_{ji} \frac{\partial}{\partial \p_i} f^K_{ij}  (\p_j - \p_i) =
- \frac{1}{k_i} \sum_j W_{ji} C^K (\p_j - \p_i) \;  ,   \label{eq-weightdynamics}
\end{equation}
where the coupling functions $C^K$ for each $K$ are the same as before. Our main result fully generalizes: if the system reaches the global minimum of $F^K_W$, then in that state there are exactly $K$ different equilibrium phase values on the nodes, separated by integer multiples of $\frac{2\pi}{K}$. The key difference is that now even in the global minimum we have $F^K_W>0$, since some links will be frustrated, as their nodes will have the same phase value (i.e. be colored the same). However, the sum of weights $W_{ij}$ of such links will be minimal, as we prove in the Appendix. Hence, the minimization of $F^K_W$ yields the improper coloring of a weighted graph using exactly $K$ colors. As in the non-weighted case, we can use either the dynamics itself or a range of other functional minimization methods.  

There are several differences with the non-weighted case. First, a graph is now $K$-colorable for any $K$, the only question is, how frustrated it will be. Our results says that the dynamics evolves to (\textit{i}) minimize the total frustration $F^K_W$, and (\textit{ii}) squeeze the remaining frustration preferentially into weaker links. We have again translated a combinatorial into a continuous optimization problem, it is just that now we cannot recognize the global minimum via $F^K_W=0$, but via $K$ different final phase values. Second, the fact that coloring is improper does not specify how many connected node pairs are colored the same (at least one), but only that the joint frustration of their links is minimal. So, we are not minimizing the number of frustrated links, but the total frustration $F^K_W$ itself. Third, in weighted case dynamics can also get stuck in the local minima. For any $K$ there is a guaranteed improper coloring corresponding to minimal total frustration, but one might again resort to ``tricks" such as special choice of initial phases or stochastic evolution to find that minimum more efficiently. On the other hand, local minima can now be interpreted as Nash equilibria in the system: let $e^K_i$ be the $K$-frustration of the node $i$ defined as the sum of $K$-frustrations of all its adjacent links $ e^K_i = \sum_j W_{ij} f^K_{ij}$. Collective dynamics can now be seen as each node $i$ seeking to minimize its $e^K_i$. Pairs of nodes connected by weak links will be more successful in this.

\begin{figure*}[!hbt]
\centering
    \includegraphics[width=\textwidth]{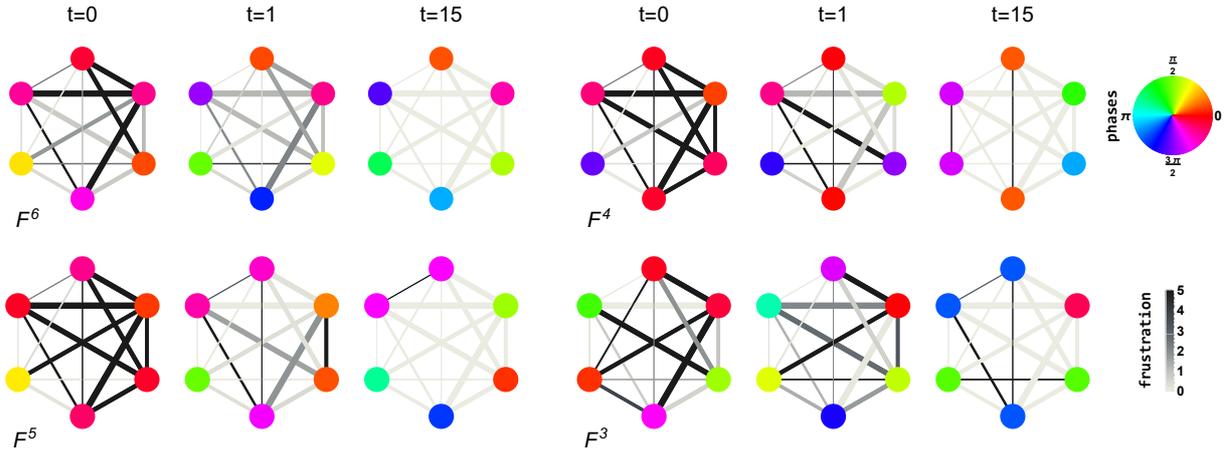} 
    \caption{Four realizations of evolution for weighted fully connected graph with $N=6$ nodes, shown in three snapshots each, as indicated. Link weights are chosen randomly between 1 to 5. Initial phases are randomly selected from von Mises distribution with $\kappa=2$. The phases of the nodes are shown by colors and frustration of links in gray-scale, as in Fig.\ref{fig-2}. One realization of evolution for each coupling function $C^6$, $C^5$, $C^4$, $C^3$ and the corresponding frustration $F^6_W$, $F^5_W$, $F^4_W$, $F^3_W$ is shown in each panel, as indicated.}
    \label{fig-4}
\end{figure*}

Next we show the performance of our dynamical method for improper coloring. For simplicity we use a fully connected graph (clique) with $N=6$ nodes (this is a non-trivial problem, a variation of the {\it channel assignment problem} for wireless graphs known from telecommunications). We randomly assign weights to its 15 links and run the dynamics of Eq.~\ref{eq-weightdynamics} with coupling functions $C^6$, $C^5$, $C^4$, $C^3$, computing the corresponding frustrations $F^6_W$, $F^5_W$, $F^4_W$, $F^3_W$. The results are shown in Fig.~\ref{fig-4}, three snapshots for each example. 
For $C^6$, as expected, the dynamics settles into a state with zero total frustration and with graph (properly) colored with $K=6$ colors. For $C^5$, as the graph tries to get colored with 5 colors, at least one pair of nodes must be colored the same. The dynamics reaches a state with just one link frustrated, the weakest one. For $C^4$, the dynamics settles into a state where two weakest links are frustrated. For $C^3$, graphs is eventually colored with 3 colors. Four links remain frustrated and we verified that their joint weight is the smallest possible. So, as expected, in all cases our dynamical coloring reached the equilibrium with minimal total frustration squeezed into weakest links. In case of a complex topology (not clique), the minimization into weakest links would interplay with topology, making it more difficult to see that algorithm works as expected.

\begin{figure*}[!hbt]
\centering
    \includegraphics[width=\textwidth]{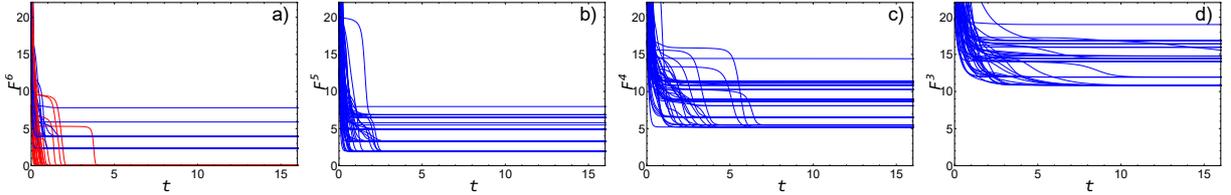} 
    \caption{(a) 50 realizations of $F^6_W (t)$ for 50 runs of Eq.~\ref{eq-weightdynamics} with $C^6$ for the weighted graph from Fig.~\ref{fig-4}. (b) The same for $C^5$ and $F^5_W$. (c) The same for $C^4$ and $F^4_W$. (d) The same for $C^3$ and $F^3_W$ (all in correspondence with four panels in Fig.~\ref{fig-4}). Plots are shown until $t=15$ for consistency. Red curves are runs that finish in zero total frustration and blue curves otherwise.}
    \label{fig-5}
\end{figure*}

To examine the impact of the local minima, we repeat in Fig.~\ref{fig-5} the analysis from Fig.~\ref{fig-3}. We track $F^K_W (t)$ between $t=0$ and $t=15$ for 50 choices of initial phases for above graph for $K=6, 5, 4, 3$, in correspondence to four panels in Fig.~\ref{fig-4}. For $K=6$ most initial phases lead to $F^6_W=0$, but not all: since the graph is weighted, despite having 6 colors available, the graph sometimes fails to ``stretch" to zero total frustration. In such equilibria, the graph is not colored by 6 colors. 
For $K=5, 4$ and 3, the global minimum that corresponds to improper coloring with $K$ colors decreases with $K$. For each of these values of $K$ there are many local minima, so the global minimum is somewhat less distinguishable, and has to be identified by actually checking if the graph is colored (although improperly) with $K$ distinct colors. We run additional simulations for larger weighted fully connected graphs and found the same scaling properties as in the non-weighted case. Global minima are increasingly harder to find as the graph size increases. Enhancements such as kicking (stochastic evolution) have the same effect as before -- they improve the search, but does not solve the problem of local minima.


\section{Discussion}
Using the analogy with repulsively coupled phase-oscillators we found a way to convert a combinatorial optimization problem (proper and improper graph coloring of non-weighted and weighted graphs) into the optimization problem of evaluating the global minimum of a real function. This function is the non-equilibrium potential for the collective dynamics of oscillators coupled through the edges of the graph (network). This dynamics can be used as a ``natural" way to search for the minimum, or alternatively, one can resort to the rich ensemble of methods from functional optimization. This conversion does not offer an immediate solution to the problem, but it allows us to approach the problem using completely different methodology. Examining in what situations this might lead to a more efficient or precise solution of the original problem remains the matter of future work, potentially extending to other discrete problems.

There are two ways of comparing our result to the state of the art. First, one could compare our collective dynamics as a method of evaluating the global minima of real functions to the existing optimization methods. We did not find this of interest, since this means comparing functional optimization methods among them, and for that there already exists abundant literature. Second and more interesting way is to examine whether solving graph coloring problem this way would be better than solving it via standard combinatorial algorithms. We did look into this, and as already stated, for all examined graphs we found the same chromatic numbers. In no case we found the chromatic number bigger or smaller than indicated by traditional algorithms. We checked this for 30 graphs with sized up to $N=100$. Another independent way of making this comparison is to look at the computational cost involved in two cases. However, this comparison strongly depends on the computing architecture, since the algorithms are of quite different nature (permutations of color arrangements vs. integrating a differential equation using e.g. Runge-Kutta integrator). An interesting direction of future work might be to find a computationally cheaper integrator for Eq.~\ref{eq-Kdynamics}~\cite{linda}.

A different issue revolves around scaling up the problem: how about coloring a graph of size $N=1000$? This represents a challenge for both combinatorial and our method. While the former suffer from combinatorial explosion, our method comes down to finding a minimum of 1000-dimensional real function, which is far from trivial. However, while cases of small graphs can in practice be easily treated by any approach, the true challenge lies in coloring large graphs, which is where our approach might make a difference. This remains the core direction of future work. Still, the degree of trigonometric polynomial in Eq.~\ref{eq-Kfrustation} is $K-1$. Hence, checking the colorability for small $K$ is relatively easy, since low-level trigonometric polynomial is not very ``wavy" and the minima is easier to find, even for large graphs. This might be a competitive advantage of our methods, since checking for these minima could beat the existing combinatorial optimization approaches. In contrast, when checking the colorability for high values of $K$, the degree of trigonometric polynomial is higher, which makes the function more wavy and hence the basins of attraction for local and global minima are steeper and harder to find. On the other hand, given the purely trigonometric nature of any $F^K$, it would be interesting to do frequency analysis on it.

We close the paper with another potentially interesting idea. Consider a graph $G$ with unknown chromatic number $\chi(G)=K_0$ and some very large tentative $K$, much larger than $K_0$ (say $K=N$). $F^K$ is zero on a potentially very large domain within $\mathbb{R}^N$, since the graph is easily $K$-colorable for this $K \gg K_0$. We call this domain support of $F^K$. Consider now $K-1$, where the support for $F^{K-1}$ is a large domain, although presumably smaller than support of $F^K$. Continuing on, one expects that support of $F^{K-2}$ is even smaller, but larger than the support of $F^{K-3}$, and so on. Eventually, the support of $F^{K_0}$ is the smallest such non-empty domain, since the support of $F^{K_0-1}$ is empty. What is the geometric relationship between these domains? For example, could they be nested subsets of each other? If so, one could be able to extrapolate $K_0$ by examining the geometric process of how these domains (supports) shrink as $K$ deceases. The value of $K$ just before this domain becomes empty is the chromatic number $K_0$, obtained without optimization.


\paragraph{Acknowledgements.} Work supported by the Slovenian Research Agency (ARRS) via programs P1-0383 and P2-0256, projects J5-8236, J1-8155, N1-0057 and N1-0071, and by EU via Marie Sklodowska-Curie Grant project 642563 (COSMOS). Part of it has been completed during the STSM of the third author (VJ) supported by the COST action CA15140. Thanks to colleagues Alexander Yurievich Gornov, Peter Koro\v sec, Ljup\v co Todorovski, and Riste \v Skrekovski for very useful suggestions and feedback.

\bigskip
\bigskip
\appendix
\section{Full proofs for the weighted and non-weighted cases}
In the appendices we report more complete and rigorous proofs for the statements made in the main text. We being by the simpler non-weighted case by noting that $K$-frustration along the link $(i,j)$ is defined as
$$
f_{ij}^K (\varphi_i,\varphi_j) = p_K(\varphi_i - \varphi_j),
$$
where $p_K$ is a trigonometric polynomial:
\renewcommand\theequation{A1}
\begin{equation}
\label{trigpol}
p_K (x) = 1 + c_K (K \cos x + (K-1) \cos 2x + \cdots + 2 \cos (K-1) x + \cos K x),
\end{equation}
where the constants $c_K$ are chosen in such a way to have $p_K(\frac{2 \pi}{K+1})=0$ satisfied.

Notice several properties of polynomials $p_K$:

P1. For each $K$, $p_K$ is a trigonometric polynomial of the degree $K$.

P2. $p_K$ are even functions.

P3. $p_K$ are $2 \pi$-periodic functions.

P4. Functions $p_K$ are nonnegative, i.e. $p_K(x) \geq 0$ for each $x$.

P5. For each $K$ the function $p_K$ has exactly $K$ local minima on $[0,2 \pi]$. These local minima are located at points $\frac{2 \pi}{m+1}$ for $m = 1,\dots,K$. (It is important to notice that there are no local minima at zero.)

{\bf Definition.}
We say that phases $\varphi_1,\dots,\varphi_N$ are in {\it $k$-regular} configuration for some $k \leq N$, if each $\varphi_i$ coincides with one of the points $z + \frac{2 s \pi}{k}$, for some $z \in [0,2 \pi]$ and $s=0,\dots,k-1$.

\medskip

Now, we are ready to several assertions that create the basis for our method of graph coloring.

\bigskip

{\bf Proposition 1.}

1. For any graph $\Gamma$ and for all integers $K$ the functions $F^K$ are nonnegative.

2. There exists an integer $m \leq N-1$, such that the value of $F^m$ at its global minimum equals zero.

3. Suppose that $F^m(\varphi_1,\dots,\varphi_N) = 0$. Then, the configuration of phases $\varphi_1,\dots,\varphi_N$ is $m+1$-regular.

\medskip

{\bf Proof.}

1. Obviously, for each $K$ the function $F^K$ is nonnegative, as it is the sum of nonnegative functions (\ref{trigpol}), see the property P4.

2. Due to property P5, polynomials $p_{N-1}$ have precisely $N$ local minima at points $\frac{2 \pi}{m}, \, m=1,\dots,N$. Let $\varphi_1, \varphi_2 = \varphi_1 + \frac{2 \pi}{N}, \varphi_3 = \varphi_2 + \frac{2 \pi}{N}, \dots, \varphi_N = \varphi_{N-1} + \frac{2 \pi}{N}$. Then, for each pair $(i,j)$, we have $f_{(i,j)}^{N-1}(\varphi_i,\varphi_j) = p_{N-1}(\varphi_i - \varphi_j) = 0$. We have $F(\varphi_1,\varphi_2,\dots,\varphi_N) = 0$. Hence, the global minimum of $F^{N-1}$ is zero. In other words, the function $F^{N-1}$ equals zero at $N$-regular configuration.

3. Suppose that $F^{m}(\varphi_1,\dots,\varphi_N) = 0$. Then all terms in the function $F^m$ are equal to zero, i.e. for $\forall (i,j) \in E, \, f_{(i,j)}^m(\varphi_i,\varphi_j) = p_m(\varphi_i - \varphi_j) = 0$. Then, from the property P5 it follows that $\varphi_i - \varphi_j = \frac{2 s \pi}{m+1}$, for some $s=0,1,\dots,m$.

\bigskip

{\bf Proposition 2.}

A proper coloring of a graph $\Gamma$ with $m$ colors exists if and only if the global minimum of $F^{m-1}$ equals zero.

\medskip

{\bf Proof.}

Suppose that for some $\varphi_1,\dots,\varphi_N, \, F^{m-1}(\varphi_1,\dots,\varphi_N) = 0$. Then, for each $(i,j) \in E$, we have that $f^{m-1}(\varphi_i,\varphi_j) = p_{m-1}(\varphi_i - \varphi_j) = 0$. Now, from the property P5 we have that $\varphi_i - \varphi_j = \frac{2 s \pi}{m}$ for some $s = 0,1,\dots,m-1$. This means that points $e^{i \varphi_1},\dots,e^{i \varphi_N}$ are located at $m$ points on the unit circle, where an angle between any two consecutive points equals $\frac{2 \pi}{m}$. Moreover, if $(i,j) \in E$, then $\varphi_i \neq \varphi_j$. Now, we assign a different color to each of $m$ points on the unit circle. Then, if $(i,j) \in E$, the corresponding nodes $q_i$ and $q_j$ are colored in different colors.

On the other side, suppose that graph $\Gamma$ can be properly colored using $m$ colors. Pick $m$ points on the unit circle, so that an angle between any two consecutive is $\frac{2 \pi}{m-1}$ and assign one of $k$ colors to each point. To each node $q_i$ assign a value $\varphi_i$, such that the point $e^{i \varphi_i}$ is colored into the same color as $q_i$. Suppose that $(i,j) \in E$. Then nodes $q_i$ and $q_j$ are colored into different colors and, hence, $\varphi_i - \varphi_j = \frac{2 s \pi}{m}$ for some $s = 1,\dots,m-1$. Then, $f^{m-1}_{(i,j)}(\varphi_i,varphi_j) = 0$. In other words, $m-1$-frustration along each link $(i,j) \in E$ is zero, and, hence, the total $m-1$-frustration $F^{m-1}(\varphi_1,\dots,\varphi_N) = 0$.

\medskip

From Proposition 2 we immediately obtain the following

{\bf Corollary.}

The chromatic number of graph $\Gamma$ is a minimal integer $m$, such that the global minimum of $F^{m-1}$ equals zero.

\bigskip

Finally, we extend above proofs to the weighted case. To do that, we prove the Proposition that guarantees that our methods for minimization of functions $F_W^{m-1}$ will end up at $m$-regular configurations. This means that minimization method for $F^{m-1}_W$ will yield a certain coloring of the graph using $m$ colors.

\bigskip

{\bf Proposition 3.}
Local minima of the function $F_W^{m-1}$ are achieved at $m$-regular configurations.

This Proposition can be easily obtained by differentiating the function $F_W^{m-1}$ and equating its derivative to zero.


\bigskip
\bigskip
\begin{thebibliography}{99}
\bibitem{nemhauser} L. A. Wolsey and G. L. Nemhauser, \textit{Integer and combinatorial optimization} (John Wiley \& Sons, 2014).
\bibitem{papadimitriou} C. H. Papadimitriou and Kenneth Steiglitz, \textit{Combinatorial Optimization: Algorithms and Complexity} (Courier Corporation, 2013).
\bibitem{schrijver} A. Schrijver, \textit{Combinatorial optimization: polyhedra and efficiency} (Springer Science \& Business Media, Vol. 24, 2003).
\bibitem{Baghel2012} M. Baghel, S. Agrawal,  S.  Silakari. "Survey of metaheuristic algorithms for combinatorial optimization." International Journal of Computer Applications 58, 19 (2012).
\bibitem{korte} B. Korte and J. Vygen, \textit{Combinatorial Optimization: Theory and Algorithms} (Springer Science \& Business Media, 2007).
\bibitem{woeginger2003} G. J. Woeginger, "Exact algorithms for NP-hard problems: A survey," in \textit{Combinatorial Optimization — Eureka, You Shrink!}, edited
by M. J{\"u}nger, et al. (Springer, Berlin Heidelberg, 2003), pp. 185-207.
\bibitem{sanz2016} S. Salcedo-Sanz, "Modern meta-heuristics based on nonlinear physics processes: A review of models and design procedures," Phys. Rep. 655, 1-70 (2016).
\bibitem{farhi2001} E. Farhi, J. Goldstone, S. Gutmann, J. Lapan, A. Lundgren, and D. Preda, "A quantum adiabatic evolution algorithm applied to random instances of an NP-complete problem," Science 292, 472-475 (2001).
\bibitem{albash2018} T. Albash and D. A. Lidar, "Adiabatic quantum computation," Rev. Mod. Phys. 90, 015002 (2018).
\bibitem{johnson2011} M. W. Johnson, et al., "Quantum annealing with manufactured spins", Nature 473, 194–198 (2011).
\bibitem{somma2008} R. D. Somma, S. Boixo, H. Barnum, and E. Knill, "Quantum simulations of classical annealing processes," Phys. Rev. Lett. 101, 130504 (2008).
\bibitem{wang2018} Z. Wang, S. Hadfield, Z. Jiang, and E. G. Rieffel, "Quantum approximate optimization algorithm for MaxCut: A fermionic view," Phys. Rev. A 97, 022304 (2018). 
\bibitem{gaitan2012} F. Gaitan and L. Clark, "Ramsey numbers and adiabatic quantum computing," Phys. Rev. Lett. 108, 010501 (2012).
\bibitem{gaitan2014} F. Gaitan and L. Clark, "Graph isomorphism and adiabatic quantum computing," Phys. Rev. A 89, 022342 (2014).
\bibitem{mcmahon2016}  P. L. McMahon, et al., "A fully programmable 100-spin coherent Ising machine with all-to-all connections," Science 354, 614-617 (2016).
\bibitem{kudo2018} K. Kudo, "Constrained quantum annealing of graph coloring," Phys. Rev. A 98, 022301 (2018).
\bibitem{arkady} A. Pikovsky, M. Rosenblum, and J. Kurths, \textit{Synchronization: A Universal Concept in Nonlinear Sciences} (Cambridge University Press, 2003), Vol. 12.
\bibitem{arenas2008} A. Arenas, A. D{\'i}az-Guilera, J. Kurths, Y. Moreno, and C. Zhou, "Synchronization in complex networks," Phys. Rep. 469, 93-153 (2008).
\bibitem{porter}  M. A. Porter and J. P. Gleeson, “Dynamical systems on networks: A tutorial,” \textit{Frontiers in Applied Dynamical Systems: Reviews and Tutorials} (Springer, 2016), Vol. 4.
\bibitem{levnajic2010} Z. Levnaji\' c and A. Pikovsky, "Phase Resetting of Collective Rhythm in Ensembles of Oscillators," Phys. Rev. E 82, 056202 (2010). 
\bibitem{luciano2011}  L. da F. Costa, et al., "Analyzing and modeling real-world phenomena with complex networks: a survey of applications," Advances in Physics 60, 329-412 (2011).
\bibitem{jacimovic2018} V. Ja\' cimovi\' c and A. Crnki\' c, "Modelling mean fields in networks of coupled oscillators," J. Geom. Phys. 124, 241-248 (2018).
\bibitem{anna2020} A. Zakharova, “Chimera Patterns in Networks: Interplay between Dynamics, Structure, Noise, and Delay,” in \textit{Understanding Complex Systems}, edited by S. Kelso (Springer, 2020).
\bibitem{acebron2005} J.A. Acebr{\'o}n, L.L. Bonilla, C.J.P. Vicente, F. Ritort, and R. Spigler, "The Kuramoto model: A simple paradigm for synchronization phenomena," Rev. Mod. Phys. 77, 137 (2005).
\bibitem{rodrigues2016} F.A. Rodrigues, T.K.D. Peron, P. Ji, and J. Kurths, "The Kuramoto model in complex networks," Phys. Rep. 610, 1-98 (2016).
\bibitem{crnkic2017} A. Crnki\' c and V. Ja\' cimovi\' c, "Exploring complex networks by detecting collective dynamics of Kuramoto oscillators," in \textit{ Proceedings of the OPTIMA-2017 Conference}, edited by Yu. G. Evtushenko, et al. (CEUR-WS, 2017), pp. 146-151.
\bibitem{zankoc2019} C. Zankoc, D. Fanelli, F. Ginelli, and R. Livi, "Desynchronization and pattern formation in a noisy feed-forward oscillator network," Phys. Rev. E 99, 012303 (2019).  
\bibitem{pietras2019} B. Pietras, N. Deschle, and A. Daffertshofer, "First-order phase transitions in the Kuramoto model with compact bimodal frequency distributions," Phys. Rev. E 98, 062219 (2019). 
\bibitem{levnajic2011} Z. Levnaji\' c, "Emergent multistability and frustration in phase-repulsive networks of oscillators," Phys. Rev. E 84, 016231 (2011).
\bibitem{levnajic2012} Z. Levnaji\' c, "Evolutionary design of non-frustrated networks of phase-repulsive oscillators," Scientific Reports 2, 967 (2012).
\bibitem{goldstein2015} D. Goldstein, M. Giver, and B. Chakraborty, "Synchronization patterns in geometrically frustrated rings of relaxation oscillators," Chaos 25, 123109 (2015).
\bibitem{astakhov2016} S. Astakhov, A. Gulai, N. Fujiwara, and J. Kurths, "The role of asymmetrical and repulsive coupling in the dynamics of two coupled van der Pol oscillators," Chaos 26, 023102 (2016).
\bibitem{wahwu} C. W. Wu, "Graph Coloring via Synchronization of Coupled Oscillators", IEEE Transactions on Circuits and Systems 45, 974-978, 1998.
\bibitem{wujiao} J. Wu, L. Jiao, R. Li, and W. Chen, "Clustering dynamics of nonlinear oscillator network: Application to graph coloring problem", Physica D 240, 1972-1978, 2011.
\bibitem{novikov} A. V. Novikov and E. N. Benderskaya, "Oscillatory Neural Networks Based on the Kuramoto Model for Cluster Analysis", Pattern Recognition and Image Analysis 24, 365–371, 2014. 
\bibitem{parihar} A. Parihar, N. Shukla, M. Jerry, S. Datta, and A. Raychowdhury, "Vertex coloring of graphs via phase dynamics of coupled oscillatory networks", Scientific Reports 7, 911, 2017.
\bibitem{chartrand} G. Chartrand and P. Zhang, \textit{Chromatic Graph Theory} (Chapman and Hall/CRC, 2008).
\bibitem{jensen11} T.~R.~Jensen, and B.  Toft, \emph{Graph coloring problems}, Vol. 39, John Wiley \& Sons (2011).
\bibitem{ganguli17} R.~Ganguli, and R.~Siddhartha, "A study on course timetable scheduling using graph coloring approach", International Journal of Computational and Applied Mathematics 12, 469-485 (2017).
\bibitem{aardal07} K.~I.~Aardal, et al., "Models and solution techniques for frequency assignment problems", Annals of Operations Research 153, 79-129 (2007).
\bibitem{Gvozdenovic2008} N. Gvozdenovi\'c and M. Laurent, "The operator $\Psi$ for the chromatic number of a graph", SIAM Journal on Optimization 19, 572-591 (2008).
\bibitem{Govorcin2013} J.~Govorčin, N. Gvozdenović, and J.  Povh, "New heuristics for the vertex coloring problem based on semidefinite programming", Central European Journal of Operations Research 21, 13-25 (2013).
\bibitem{Jabrayilov2018} A.~Jabrayilov, and P.~Mutzel, "New integer linear programming models for the vertex coloring problem", In Latin American Symposium on Theoretical Informatics (pp. 640-652). Springer (2018).
\bibitem{knuth1994} D. E. Knuth, "The sandwich theorem," The Electronic Journal of Combinatorics 1, 1 (1994). 
\bibitem{Eppstein2003} D. Eppstein, "Small maximal independent sets and faster exact graph coloring", J. Graph Algorithms Appl. 7, 131-140 (2003).
\bibitem{Malaguti2010} E. Malaguti and P. Toth, "A survey on vertex coloring problems", International transactions in operational research 17, 1-34 (2010).
\bibitem{lewis}  R. Lewis, \textit{A Guide to Graph Colouring: Algorithms and Applications} (Springer, 2016).
\bibitem{sundaram} R. K. Sundaram, \textit{A First Course in Optimization Theory} (Cambridge University Press, 1996).
\bibitem{mardia} K. V. Mardia and P. E. Jupp, \textit{Directional Statistics} (John Wiley \& Sons, 1999).
\bibitem{riste1999} R. \v Skrekovski, "List improper colourings of planar graphs," Combin. Probab. Comput. 8, 293-299 (1999).
\bibitem{mohar2003}  B. Mohar, "Circular colorings of edge‐weighted graphs," J. Graph Theory 43, 107-116 (2003).
\bibitem{mishra2005}  A. Mishra, S. Banerjee, and W. Arbaugh, "Weighted coloring based channel assignment for WLANs," Mobile Comput. Commun. Rev. 9, 19-31 (2005).
\bibitem{linda} U. Ascher and L. Petzold, \textit{Computer Methods for Ordinary Differential Equations and Differential-Algebraic Equations} (SIAM, 1989).
\end{thebibliography}
\end{document}